Hypothesis

**Three phases in the evolution of the standard genetic code: how translation could get started**

Peter T. S. van der Gulik
Centrum voor Wiskunde en Informatica, Amsterdam, The Netherlands.

**Abstract: A primordial genetic code is proposed, having only four codons assigned, GGC meaning glycine, GAC meaning aspartate/glutamate, GCC meaning alanine-like and GUC meaning valine-like. Pathways of ambiguity reduction enlarged the codon repertoire with CUC meaning leucine, AUC meaning isoleucine, ACC meaning threonine-like and GAG meaning glutamate. Introduction of UNN anticodons, in a next episode of code evolution in which nonsense elimination was the leading theme, introduced a family box structure superposed on the original mirror structure. Finally, growth rate was the leading theme during the remaining repertoire expansion, explaining the ordered phylogenetic pattern of aminoacyl-tRNA synthetases. The special role of natural aptamers in the process is high-lighted, and the error robustness characteristics of the code are shown to have evolved by way of a stepwise, restricted enlargement of the tRNA repertoire, instead of by an exhaustive selection process testing myriads of codes.**

**Introduction**

I wish to suggest a model for the evolution of the standard genetic code. "Explaining *how* the genetic code evolved and *why* the codon assignment took its present form" (Delarue, 2007) is, "arguably, the central and the hardest problem in the study of the origin of life, and one of the hardest in all evolutionary biology" (Wolf and Koonin, 2007). In this paper I suggest that different selection pressures during different phases in the process of evolution of the standard genetic code caused superposed patterns in the genetic code table, which is why it is so hard to read the patterns. I will argue that successive periods of ambiguity reduction, nonsense elimination and growth rate optimization each left their own mark on the genetic code table.

**A primordial 4 tRNA genetic code**

Ardell (1998) made the observation that "the GNC-encoded amino acids (glycine, alanine, aspartate and valine) are also the most abundant amino acids produced in the Miller-Urey reactions and found in the Murchison Meteorite (Miller 1987)". This observation naturally leads to the vision of a simple early genetic code encoding only 4 amino acids: glycine, alanine, aspartate and valine. Although restricted in diversity, this early genetic code would suffice to make simple proteins: valine would be the residue to build a hydrophobic core, aspartate would be the one to build a polar exterior, glycine would be part of turns, and alanine would be available to sit in between other residues wherever needed. So, the essential requirements of protein structure are served by just these 4 amino acids, which is a staggering realization. It gives the opportunity to get out of the "chicken-and-egg problem" (Delarue, 2007) of the need of having proteins present to be able to develop a translation apparatus: one can now envision *simple proteins* being present before a complex process like e.g. histidine biosynthesis is becoming part of the biochemical repertoire. This offers us the escape route out of the paradox (described by Wolf and Koonin, 2007) that proteins are a necessary part of a translation system, accurate and efficient enough to be able to produce proteins.

### *the tRNAs*

The 4 amino acid genetic code could be read by just 4 tRNAs: the tRNAs with the anticodons GCC, GGC, GUC and GAC. These anticodons are remarkably similar in structure (which is another way of saying that the 4 most abundant prebiotic amino acids are found in the upper half of the lowest row of the genetic code table), which naturally leads to the suggestion that one of them was the primordial molecule, and the other 3 developed by duplication and diversification from this primordial molecule (see Hittinger and Carroll (2007) for a background on the importance of gene duplication, diversification and subfunctionalization in cellular evolution). These tRNAs weren't necessarily restricted in function to only serve the synthesis of simple early proteins: tRNAs carrying aspartate and glycine could be implicated in adenosine synthesis in the RNA world ("The purine ring is assembled on ribose phosphate ... Glycine joins phosphoribosylamine ... *aspartate contributes only its nitrogen atom to the purine ring*" Stryer, 1988), to give an example. Amino acids as ribozyme cofactors is another, more well-known, example of a proposed tRNA function (Szatmáry, 1993), apart from protein synthesis.

### *the codons*

The simple protein-coding genes using the 4 tRNA genetic code could be written in just 4 codons: GGC, GAC, GCC and GUC. The special thing of this 4 codon genetic code is that GGC is mirrored by GCC and GAC is mirrored by GUC. This means that both template strand and coding strand could serve as mRNA. Point mutations leading to one of the 60 remaining codons would be selected against, because the resulting mRNA could not be read by the 4 present tRNAs (or the mRNA produced from the resulting RNA could not be read by the present tRNAs, in case the mutations resulted in GGU, GAU, GCU or GUU). While being bound within the restrictions of 4 codons, and within the restrictions of short peptides (because of the limits set to genome size by the error catastrofe resulting from poor replication as pointed out by Eigen and Schuster, 1977), variation of the sequences within this limitations could scan the sequence space for useful molecules. As de Duve (2005) pointed out, such explorations of limited sequence spaces can even take the form of exhaustive exploration, just because the number of combinatorial possibilities is not extremely large.

### *frozen ambiguity*

An important implication of simple early protein structure being realized by valine, aspartate and glycine on critical positions, is that the "frozenness" first pointed out by Crick ("... the code is universal because ... *any change would be lethal* ..." Crick 1968) would be present in a much earlier stage than Crick in that paper envisioned.

When a new amino acid was introduced in simple early protein sequences, the moment it took an important place in one of them (a place in which *that* amino acid, and no other, should be), the meaning of the codon was frozen, if that protein was essential. Only when the number of essential positions in which a certain codon was used, went back to zero, the codon "thawed" again, which happened a few times in evolution, especially in mitochondrial lineages (Knight et al., 2001).

Woese (1965) pointed out that early simple protein synthesis was necessarily of such a poor quality, that the resulting proteins have to have been *statistical proteins* (Woese, 1965). In this vein, I suggest that the 4 original codons were each effectively coding for a *group* of amino acids: GUC meaning valine/isoleucine/leucine, GCC meaning alanine/threonine/proline/serine, GAC meaning aspartate/glutamate, and the "group" of amino acids encoded by GGC having just one member, glycine. As long as a residue in a simple early protein just meant "polar exterior" this level of ambiguity was acceptable. But as soon as protein sequences started to interact specifically with RNA sequences, the level of ambiguity became problematic.

**Different pathways of ambiguity reduction**

*the first column*

The sequence UAUUGGGG appears time and again when RNA sequences are screened for isoleucine binding (Caporaso et al. 2005 and references therein). This fact offered primitive biochemistry a strategy to reduce ambiguity of the GUC codon. By producing such an isoleucine aptamer, the microenvironment of translation could be "scavenged" for isoleucine. As a result, the proportion of tRNA$^{Val}$ charged with valine or leucine was slightly higher than in ribocells without isoleucine aptamer. The valine specificity could be further improved by the emergence of a tRNA with anticodon GAU (emerging as a result of a point mutation in an anticodon in a duplicate of a gene for tRNA$^{Val}$). Like isoleucine, this tRNA would bind the aptamer binding-site, and in this way the local concentrations of isoleucine and tRNA$^{Ile}$ would both be higher around the isoleucine aptamer. When the primordial Class Ia aminoacyl-tRNA synthetase was in the isoleucine aptamer micro-environment, it would be charging tRNA$^{Ile}$ with isoleucine; when it was outside that micro-environment, it would be charging tRNA$^{Val}$ with valine or leucine.

The appearance of AUC codons in the mRNA sequences didn't cause the emergence of unreadable mirror molecules: the mirror codon GAU would be read by the tRNA$^{Asp}$ with anticodon GUC, according to the Wobble Rules (Crick 1966), which thus are rules, more than 3 billion years old.

The emergence of two tRNAs, one transferring only isoleucine, and the other just leucine and valine, where before just one tRNA existed, transferring all three amino acids, is an example of duplication followed by subfunctionalization (Lynch and Force, 2000).

*the second column*

Because alanine and glycine are the most simple amino acids, it is not too far fetched to assume that for those two, a charging mechanism existed that entirely consisted of RNA. Recently, a proposal was made, that such charging by ribozymes was directly coupled to production from keto acids (Copley et al., 2005). Charging tRNA$^{Ala}$ with threonine, proline or serine, would then be a novelty, introduced with the emergence of the primordial Class IIa aminoacyl-tRNA synthetase. On the one side, loss of specificity of alanine coding was a disadvantage. On the other side, gain of residues like proline or serine in proteins was an advantage. A way out of this "natural selection's dilemma" was the emergence of a new tRNA with anticodon GGU (emerging as a result of a point mutation in an anticodon in a duplicate of a gene for tRNA$^{Ala}$), combined with a change in the original tRNA$^{Ala}$, in a way that it was no longer recognized by the Class IIa aminoacyl-tRNA synthetase. When the original RNA system for production and charging of alanine was still working, this could happen without all mRNA being suddenly unreadable.

Like the case with isoleucine, this ambiguity reduction doesn't result in unreadable mirror mRNAs clogging the ribocell, because GGU (the mirror codon) is read by the tRNA$^{Gly}$ with anticodon GCC.

Again, we saw duplication followed by subfunctionalization, and we will see it also in the other pathways of ambiguity reduction.

### *the third column*

A third pathway of ambiguity reduction is seen for the GAC codon. Here, appearance of a tRNA with anticodon CUC (emerging as a result of a point mutation in an anticodon in a duplicate of a gene for tRNA$^{Asp}$) provided the opportunity to develop specialization of the GAU and GAC codons (read by the tRNA with anticodon GUC) to being unambiguous aspartate codons, and to develop specialization of the GAG codon (read by the tRNA with anticodon CUC) to being a glutamate codon. Likely, the primordial Class Ib and Class IIb aminoacyl-tRNA synthetases played a leading role in this specialization process. Steitz and Moore (2003) hinted as a possible primordial protein role being "chaperones" for ribozymes; Schimmel and Ribas de Pouplana (2001) argued that Class I and Class II molecules originally "embraced" the tRNA in a coordinated manner. Such "embracing" is exactly what is expected from chaperone-like simple proteins acting on ribozymes. The 3 molecule complex could be in a slightly different conformation when the two synthetases "embraced" tRNA$^{Glu}$ compared to when they "embraced" tRNA$^{Asp}$, and these different conformations could result in the one, or the other amino acid being charged from the one side, or the other side by the Class I or the Class II aminoacyl-tRNA synthetase.

The appearance of GAG codons in mRNA did lead to the appearance of CUC codons in the mirror mRNAs; however, this was no problem because these codons were simultaneously appearing as a result of the fourth pathway of ambiguity reduction!

### *first column part two*

This fourth pathway of ambiguity reduction concerned the leucine "pollution" of valine. The emergence of a tRNA with anticodon GAG (emerging as a result of a point mutation in an anticodon in a duplicate of a gene for tRNA$^{Val}$) could lower this "leucine pollution". GNN sequences are repeatedly found as part of aptamer binding sites for the amino acid encoded by the codon, recognized by the particular sequence as an anticodon (Caporaso et al. 2005 and references therein). This holds for GAA and phenylalanine, GUA and tyrosine, GUG and histidine, GAG and leucine. A possible explanation would be, that the anticodon loop of the tRNA itself would, in certain sequence contexts, have been working as an aptamer binding-site. In that case a tRNA with an anticodon GAG would be in the middle of a micro-environment with a slightly heightened leucine concentration. In the micro-environment of tRNA$^{Leu}$ the primordial Class Ia aminoacyl-tRNA synthetase would be charging tRNA$^{Leu}$ with leucine; outside that micro-environment it would be charging tRNA$^{Val}$ with valine (unless it was in the micro-environment of the isoleucine aptamer).

### *tRNAs and codons*

The 4 different pathways of ambiguity reduction resulted in a 8 tRNA genetic code. I briefly recapitulate what has emerged. CUC was read as leucine by a tRNA with anticodon GAG. AUC was read as isoleucine by a tRNA with anticodon GAU. GUC was read as valine by a tRNA with anticodon GAC.
ACC was read as threonine/proline/serine by a tRNA with anticodon GGU. Because the hydroxy group of serine and threonine engages in an intramolecular hydrogen bond with the backbone nitrogen atom, resembling the imino structure of proline, these three amino acids are very much "related" (an observation made by Woese et al., 2000). GCC was read as alanine by a tRNA with anticodon GGC.
GAU and GAC were read as aspartate by a tRNA with anticodon GUC. GAG was read as glutamate by a tRNA with anticodon CUC.
And the codons GGU and GGC were read as glycine by a tRNA with anticodon GCC. Although 10 codons now carried meaning, this left 54 codons unassigned. As Sonneborn (1965) first pointed out, this led to an awful lot of nonsense mutations. Because the number of protein genes, and the proportion of protein-coding sequence in the genome, had of course massively increased during the ambiguity reduction phase of the evolution of the genetic code (as part of an endless series of positive feedback loops of increasing specificity of replication and the resulting genome size increase) these nonsense mutations became a serious factor, which could be addressed by natural selection. The next phase of the evolution of the genetic code was therefore characterized by nonsense elimination.

### **No Nonsense!**

The 4 tRNA genetic code had only tRNA anticodons of the structure GNC. The process leading to a 8 tRNA genetic code had expanded anticodon structure to GAG, GAU, GGU and CUC. The next anticodon structure expansion was to have tRNAs emerging with anticodons starting with unmodified U. In the first position of the anticodon, unmodified U recognizes all 4 possibilities in the third position of the codon: U, C, A and G (see e.g. Takai and Yokoyama, 2003). Having tRNAs appearing with anticodons UCC, UGU, UGC, UAG, UAU and UAC (as variants of tRNAs with respectively anticodons GCC, GGU, GGC, GAG, GAU and GAC) reduced the potential for nonsense mutation significantly (of course this strategy could not be followed for the GAN box). The family box structure of the genetic code table was the result of the introduction of unmodified U in the first position of the anticodon of tRNAs.
Another result was the destruction of the original mirror structure of the genetic code. But this structure had become cumbersome anyway. Proteins had become so large and their sequences so specific, that it had become "nonsensical" to produce their mirror proteins too. The way to get rid of that system was: having a Shine-Dalgarno sequence upstream from the gene, distinguishing coding strand and template strand by that sequence, and having a 16S rRNA doing that job.

The ultimate result is: having a genetic code table with a family box pattern superposed on a more primitive mirror codons pattern. *That* was a mess to sort out! Especially because in the third phase presented in this paper (of course this paper ignores the really early phases) a third pattern was superposed on the first two...

**An enforced order of subsequent assignment for the remaining codons**

Ardell (1998) made the observation that "an in vitro-evolved telomerase-like ribozyme was polymerizing nucleotides 10-40 times more efficiently and with a higher fidelity when directed by G and C template residues than with A and U residues (Ekland and Bartel 1996)". This observation naturally leads to the realization that ribocells grew faster when they were GC-rich, if this phenomenon is a general one for RNA-polymerizing ribozymes. This is not a far fetched assumption: of course it is more easy to polymerize nucleotides with 3 hydrogen bonds than it is to polymerize nucleotides with 2 hydrogen bonds!
With problems in ambiguity and nonsense mutation strongly reduced, the next factor guiding genetic code expansion was growth rate. If we assume that it's not just GC-richness which influences RNA-polymerization, but that also the specific neighbours of a nucleotide influence the speed of polymerization on that stretch of RNA, we can propose the existence of "fast codons" and "slow codons" in the RNA World ("fast" and "slow" with respect to their *replication*). Trifonov (2000) gives a table with the melting enthalpies of the 32 codon pairs, and this provides us with exactly the order of speed we are looking for. The "fastest codons" are the pair GGC – GCC. This means that as soon as growth rate became a serious factor in the competition of ribocells with developing genetic codes, those ribocells which used a lot of alanine and glycine in their proteins would survive. At the other hand, codon pairs which were having A and U in both the first and the second position in the codon of both members of the pair, or just one member of the pair, were slow. This means that ribocells which used scarcely any isoleucine would grow fast. Maybe the lower half of the AUN family box "thawed" again. Because a protein contains just one start codon, AUN was a perfect candidate to work in concert with the Shine-Dalgarno sequence.

*Proline, Serine*

Which pair of codons is the fastest when we exclude GGC – GCC? It is GGG – CCC. GGG was already in use as part of the glycine family box. But CCC was still unassigned. By assigning CCC to proline, biochemistry was establishing three results in just one move: further ambiguity reduction (and of course, from the viewpoint of protein structure it is urgent to be able to encode proline unambiguously!), further nonsense elimination (the appearance of a tRNA$^{Pro}$ with anticodon UGG further expanded the family box structure of the genetic code) and further growth rate optimization (the more proline containing turns in the proteins, the faster the ribocell grows) (do we still call this a ribocell?).
The primordial Class IIa aminoacyl-tRNA synthetase was still not powerful enough to distinguish proline and serine. But ambiguity reduction isn't only produced by more specific charging of tRNA. It can also be produced by selective destruction of mis-charged tRNAs! Zhu et al. (2007) describe the action of independent proteins which destroy mis-charged tRNAs.

Which pair of codons is the fastest when we exclude both GGC – GCC and GGG – CCC? GGA – UCC. Again producing further ambiguity reduction (unambiguous serine coding), further nonsense elimination (another family box) and further growth rate optimization (the more serine, proline, alanine and glycine, the faster the growth).

*Arginine, Cysteine*

Which pair of codons is fastest when we exclude the first 3 pairs? Three pairs of codons do have nearly the same melting enthalpy: GCG – CGC, GCU – AGC and GCA - UGC. CGC became arginine! Again producing further nonsense elimination (one more family box) and further growth rate optimization. Arginine thus emerges as one of the five "fastest" amino acids. Natural selection pressed the cell to use as much arginine as possible (together with glycine, alanine, proline and serine). Arginine is not a prebiotic amino acid. We now move out of the prebiotic world and into the biosynthetisized world. And which "biotic" amino acid is handled first? The one biochemistry can manipulate with an aptamer. Arginine sticks to CGU, CGC, CGA and AGG sequences in aptamer binding-sites. The tRNAs with anticodons GCG and UCG (and later CCA) also stick to these aptamers. The new Class Ia aminoacyl-tRNA synthetase which dealt with these tRNAs did find them around (and on) the arginine aptamer. This new ArgRS was coevolving with a new Class IIa aminoacyl-tRNA synthetase, AlaRS, which at this stage took over charging $tRNA^{Ala}$ from the original ribozymatic mechanisms. For an overview of coevolving Class I and Class II aminoacyl-tRNA synthetases, and the striking mirror pattern of the Class I phylogenetic tree versus the Class II phylogenetic tree, see Schimmel and Ribas de Pouplana (2001); as a counterpart to TrpRS, SepRS has recently been discovered (O'Donoghue et al., 2005; Hohn et al., 2006). Phosphoserine (Sep) is the precursor of cysteine: in some archaea $tRNA^{Cys}$ is charged with phosphoserine, which is subsequently, while being covalently bound to $tRNA^{Cys}$, enzymatically processed to become cysteine (Yuan et al., 2006).

AGC became serine, like UCC. Apparently the property to unambiguously encode serine developed twice: an early example of convergent evolution. No family box this time. A tRNA with anticodon UCU would stick to an arginine aptamer...

UGC became cysteine. That amino acid gave biochemistry a *huge* leap forward. Note that the $tRNA^{Ser}$ with anticodon GCU is an excellent candidate to lead, by duplication and diversification, to the emergence of a $tRNA^{Cys}$ with anticodon GCA. But note also the presence of Class Ic and Class IIc aminoacyl-tRNA synthetases: coevolving TrpRS and SepRS. What is tryptophan doing so early in the model? Well, tryptophan is a biosynthetic precursor of NAD ("*Nicotinate* [...] is derived from tryptophan" Stryer, 1988), a relict from the RNA World, so it is not strange to find an aptamer (tryptophan sticks to binding sites containing CCA triplets) manipulating tryptophan. As a molecule handled by a tRNA, tryptophan is already present, as a protein constituent it will appear in the model in the next step. The $tRNA^{Trp}$ thus originally had a different function apart from protein synthesis. We see that besides (and next to) the possibility of tRNAs having been "coding coenzyme handles" (Szatmáry, 1993), there is also the possibility of tRNAs having been "coding metabolite handles".
This also focuses on the existence of SepRS, prior to the emergence of CysRS. Wong and Di Giulio have repeatedly summoned attention for the role of tRNA as a platform for amino acid biosynthesis (Wong, 1975, Wong, 2007 and references therein, Di Giulio 2005 and references therein). This point should not be taken too far, in the sense that this way of synthesis would also hold for prebiotic amino acids, or that no other patterns would be present in the code. But providing indeed obviously the more primitive way to produce Cys-tRNA$^{Cys}$, SepRS had to exist prior to the emergence of CysRS. We thus can envision a stage in the evolution of the genetic code with at least 14 different tRNAs (tRNA$^{Leu}$, tRNA$^{Ile}$, tRNA$^{Val}$, two kinds of tRNA$^{Ser}$, tRNA$^{Pro}$, tRNA$^{Thr}$, tRNA$^{Ala}$, tRNA$^{Asp}$, tRNA$^{Glu}$, tRNA$^{Cys}$, tRNA$^{Trp}$, tRNA$^{Arg}$ and tRNA$^{Gly}$), but with only 4 couples of aminoacyl-tRNA synthetases charging 13 of these tRNAs: the original Class Ia – Class IIa couple, the ArgRS – AlaRS couple, the GluRS – AspRS couple, and the TrpRS – SepRS couple.

Later, a new CysRS would coevolve with a new GlyRS, taking over charging tRNA$^{Gly}$ from the original ribozymatic mechanisms.

### Tryptophan

Which pair is next? GGU – ACC. Glycine and threonine. Already present from the earlier phases of genetic code building.
Which pair is next? Four pairs do have nearly the same melting enthalpy: CCG – CGG (already present as members of family boxes), CCU – AGG (more tRNA sticking to arginine aptamer), CCA – UGG (tryptophan entering protein sequences) and GAC – GUC (these were already frozen, in an ambiguous form, in the 4 tRNA genetic code).

### Stop

Which pair is next? Again four pairs do have nearly the same melting enthalpy: UCG – CGA, UCU – AGA, UCA – UGA and GAG – CUC. Now we are thoroughly in the Protein World. To recognize AGA (without recognizing serine codons), modification of U is necessary (see Numata et al., 2006). And UGA is recognized by a Release Factor! Recently a hypothesis was proposed that release factors were a strategy to cope with frame-shift mutations (Itzkovitz and Alon, 2007)!

### Histidine, Glutamine

Which pair is next? Five pairs do have nearly the same melting enthalpy: ACG – CGU, ACU – AGU, ACA – UGU, GAU – AUC and CAC – GUG. Histidine sticks to the GUG anticodon.
Which pair is next? CAG – CUG. A tRNA$^{Glu}$ with anticodon CUC evolves towards a tRNA$^{Gln}$ with anticodon CUG, another case of "pretran synthesis" (Wong, 2007). It is highly relevant to note here, that an amino acid with a polar requirement (Woese et al., 1966) of 12.5, is modified to become an amino acid with a polar requirement of 8.6, while the amino acid already encoded by CAC (histidine) has a polar requirement of 8.4. Histidine did become encoded by that codon because of chemical (aptamer) reasons; this is not the case for glutamine. *Glutamate evolved in a direction to match histidine's polar requirement*. This means that the selective pressure is higher than just protein structure can account for. For this reason, the concept that certain critical residues in proteins *had* to stay the same polar requirement, because of essential interactions with molecules like NAD, FAD, FMN, THF, and so on, is a fundamental part of the model presented in this paper.

*Methionine*

Which pair is next? CAU – AUG. Coevolution of MetRS and HisRS is seen in the mirror pattern of Class I aminoacyl-tRNA synthetase and Class II aminoacyl-tRNA synthetase phylogeny (Schimmel and Ribas de Pouplana, 2001). MetRS could already have been around for some time, before AUG did appear in protein coding sequences, because it was producing and handling methionine as part of methylation biochemistry.
*We again see, that a new amino acid (methionine, with a polar requirement of 5.3) is ultimately favored to be assigned to a codon, because it matches the previously assigned amino acid (isoleucine, with a polar requirement of 4.9) of the box in polar requirement.*

*Phenylalanine, Tyrosine, Asparagine, Lysine*

Which pair is next? Two pairs do have nearly the same melting enthalpy: GAA – UUC (phenylalanine sticks to GAA in the anticodon) and UAC – GUA (tyrosine sticks to GUA in the anticodon). Coevolution of a new couple of Class Ic and Class IIc aminoacyl-tRNA synthetases: TyrRS and PheRS.
Which pair is next? Again two pairs: UAG – CUA and AAC – GUU. Another stop codon and a tRNA$^{Asp}$ with anticodon GUC evolving towards a tRNA$^{Asn}$ with anticodon GUU (the "pretran synthesis" pathway again: Wong, 2007).
Which pair is next? Two pairs again: AAG – CUU and CAA – UUG. A new couple of Class Ib and Class IIb aminoacyl-tRNA synthetase (LysRS-I and LysRS-II) and leucine sticking to CAA in the anticodon of a tRNA. *We again see, that lysine (polar requirement 10.1) has evolved towards the polar requirement of previously assigned asparagine (polar requirement 10.0), lysine being obviously derived from arginine (polar requirement 9.1). We also see that chemically determined leucine (polar requirement 4.9) is fully acceptable for matching previously assigned (and also chemically determined) phenylalanine (polar requirement 5.0), a constellation ultimately determined by the preadaptive assignment of GUC to "valine-like". Another preadaptive assignment in a step already discussed, is UGC being assigned to cysteine (polar requirement 4.8 or 5.5), when the later in protein sequences appearing tryptophan (polar requirement 5.2 or 5.3) is chemically determined to be encoded by CCA (the uncertainties in polar requirement in the UGN box are due to discrepancies between Woese et al., 1966 and Woese, 1973).*

The slowest codons are those of the 4 remaining pairs: the codons which are not having any G or C at all. One of them is taken by Release Factor, the other 7 start to encode amino acids already part of the repertoire.

**Perspective**

Now I bring the model proposed in this paper in perspective with some other ideas brought forward in the field. In many cases, the model is not in conflict with these ideas, but results in a coherent picture, in combination with them.
As an example, consider the ideas of Vetsigian et al. (2006). Massive horizontal gene flow should have happened during early stages of evolution. The model presented in this paper is not in conflict with that model. The new tRNAs, appearing as the result of ambiguity reduction, nonsense elimination and growth rate optimization, could very well have spread through the RNA World by means of horizontal gene flow, as proposed by Vetsigian et al. (2006). However, the model is also compatible with more traditional ideas of vertical gene flow and competition between distinct cell lines. The way the new tRNA genes spread and the order in which the new tRNA genes emerged are in fact two independent problems.

*Wolf and Koonin*

In the same way, the model proposed in this paper is both compatible with the panorama of Wolf and Koonin (2007) that the large rRNA was running peptide bond formation right from the start, and with the panorama of Woese (2001) that originally, mRNAs and tRNAs were running protein synthesis on their own (cyclic conformational change of tRNAs effecting both peptide bond formation and mRNA movement) and rRNAs emerging only later, when genomes were evolving away from the "Eigen Cliff" (vividly drawn in Wolf and Koonin, 2007). The way the peptide bond was originally forged, and the order in which the new tRNA genes emerged are independent problems too.

The model proposed in this paper (the AR-NE-GRERO model: Ambiguity Reduction – Nonsense Elimination – Growth Rate and Error Robustness Optimization model) is not compatible with the elements of the model of Wolf and Koonin (2007) that mRNA was a late development, and that dipeptides and small oligopeptides of complete arbitrary amino acid composition and complete arbitrary amino acid sequence were the original product of prototranslation. The possibility to genetically hold on to encode "hydrophobic", "polar", "turn" and "neutral" on critical positions in short peptides is the key to get out of the paradox (sharply described by Wolf and Koonin, 2007). The flaw in their argument seems to be the concept that amino acids like histidine or tyrosine are essential for making protoproteins right from the start.

Wolf and Koonin (2007) state that "the catalytic domains of the Class I aaRS form but a small twig in the evolutionary tree of the Rossman fold proteins; the advent of the common ancestor of the aaRS is preceded by a number of nodes along the evolutionary path". In the AR-NE-GRERO model, aminoacyl-tRNA synthetases appear on the scene very early, and coevolve with the genetic code. There is no room for a family of ribozymes performing the original tRNA aminoacylation, except for the very earliest amino acids, like alanine and glycine. And there is also no room for a wide variety of Rossman fold proteins performing all kind of different biochemistry.

However, that doesn't necessarily makes the AR-NE-GRERO model useless. Ibba and Francklyn (2004) describe how a GluRS evolves into a protein glutamylating a substrate different from the CCA terminus of a tRNA. It is logical that primitive biochemistry is using the means which it has, to grow and develop, that means: using aminoacyl-tRNA synthetases to develop new functions, outside protein synthesis. When biological sequences stride into new territory, sequence change can suddenly switch to a much higher tempo. Phylogeny has been more than once the victim of these kind of artefacts. I suggest that the nodes mentioned by Wolf and Koonin in reality come behind the diversification from the aminoacyl-tRNA synthetases, and that these Rossman fold proteins have evolved from particular aminoacyl-tRNA synthetases, but have lost common characteristics, and are no longer recognizable as descendants of particular aminoacyl-tRNA synthetases.

## *Delarue*

The AR-NE-GRERO model is *not* compatible with the model proposed by Delarue (2007), the "Binary Differentiation Model". The first codons to be unambiguously coding are GGC, GCC, GAC, GUC, GAG, CUC and AUC in the AR-NE-GRERO model. In the Binary Differentiation Model, the first step is having two kinds of assignment, all codons of the left half of the table being "amino acid codons" and all codons at the right half of the table being "stop codons". Both groups are then subdivided: the left half being subdivided in what one could call "P.R. 5 amino acids" (first column) and "P.R. 7 amino acids" (second column), and the right half being subdivided in "Stop Codons" (third column) and what one could call "P.R. 8 amino acids" (fourth column). Although there are certain parallels with the AR-NE-GRERO model (when one assumes that the first 4 tRNAs could suppress nonsense mutations concerning their own column, the 4 column structure emerges), the presence of stop codons right from the start and the rigid meaning-switching mechanism of subdivision are alien to the AR-NE-GRERO model. In fact I feel that the rigid meaning-switching mechanism is *enforced* on the real genetic code table, by assigning two codons each to methionine and tryptophan, denying AGA and AGG their arginine identity, and denying PheRS its Class II identity.

## *Wong, and Di Giulio*

The AR-NE-GRERO model differs from the ideas of Wong (1975; 2007, and references therein), and of Di Giulio (2005, and references therein) mainly in that these authors start their scenarios of genetic code evolution with all 64 codons carrying meaning (like Delarue, 2007). This approach is based on the thorough influence of Sonneborn (1965) and Crick (1968) on the field. I believe it is wrong to propose scenarios of "codon take-over", it is in conflict with the principle of continuity and the primitive proteins are the key to get out of the paradox of impossible protein-less translation (described by Wolf and Koonin, 2007, and, as "the chicken-and-egg problem", by Delarue, 2007).
Part of the AR-NE-GRERO model is perfectly compatible with the ideas of Wong and Di Giulio. When Wong (2007) writes: "The coevolution theory proposes that primordial proteins consisted only of those amino acids readily obtainable from the prebiotic environment, representing about half the twenty amino acids of today, and the missing amino acids entered the system as the code expanded along with pathways of amino acid biosynthesis", that is exactly what the AR-NE-GRERO model shows. Also the concept of biosynthesis of a new amino acid on the tRNA fits perfectly in the AR-NE-GRERO model, as long as it is restricted to *some* of the amino acids, like cysteine or glutamine. For some other amino acids, like phenylalanine or histidine, the AR-NE-GRERO model envisions a different route of recruiting: aptamer binding-sites.
When the route of introduction of a new amino acid was synthesis on the tRNA, this by no means excludes selection for error robustness. With glutamine, processing of Glu-tRNA$^{Gln}$ to Gln-tRNA$^{Gln}$ was the way of developing glutamine, but the polar requirement of glutamine (8.6) which resembled that of histidine (8.4) so much, was the factor that decided that the assignment of CAR to glutamine was the one emerging out of the competition.

*Taylor and Coates*

Often the grouping of the amino acids in rows in the genetic code table is said to be a grouping according to biosynthetic pathway (Taylor and Coates, 1989; Freeland et al., 2000 and references therein). I strongly disagree with putting leucine and histidine in one such group with proline, glutamine and arginine. I also disagree with denying AGN codons their serine and arginine identity. And I argue that part of the "biosynthetic pattern" of rows is caused by chance. Isoleucine is in the third row because of the chemical fact of UAUUGGGG binding isoleucine, not because aspartate is a biosynthetic precursor of isoleucine. When tRNA$^{Asp}$ with anticodon GUC has the choice to point mutate to either GUA, GUG or GUU, there is a chance that asparagine will end up in the row of isoleucine, the more so because GUA and GUG cannot be chosen because of chemical determination for these anticodons to be assigned to decoding tyrosine and histidine.

*Sella and Ardell*

Sella and Ardell summon attention to the coevolution of genetic codes and genes (Sella and Ardell, 2006, and references therein). They argue that redundancy and error robustness are bound to emerge, in the coevolutionary processes of expansion of the number of genes and expansion of the diversity of coded amino acids. Such a process is exactly described in the AR-NE-GRERO model. While Sella and Ardell describe the general dynamics of evolving genetic codes, the AR-NE-GRERO model describes one particular such process, namely the one which led to the standard genetic code.
The only real difference between the AR-NE-GRERO model and the ideas of Sella and Ardell is that they start out with 64 ambiguous codons (like Wong, Di Giulio, and many others) where the AR-NE-GRERO model starts with nearly all codons unassigned.

*Trifonov*

The basic process of evolving new codons and new tRNAs, which is at the heart of the AR-NE-GRERO model, was published in Trifonov (2000). The elements of codon capture as used in Trifonov's model, and of the method to derive "consensus temporal order" are alien to the AR-NE-GRERO model. Codon capture (see e.g. Jukes, 1981; Osawa and Jukes, 1989; Osawa, 1995; Knight et al., 2001) is in my view a phenomenon of the DNA World, not a phenomenon of the RNA world.

*Knight, Freeland and Landweber*

Of course, the AR-NE-GRERO model has elements taken from Woese (1965), Crick (1968), Knight et al. (1999) and de Duve (2005). The concept of statistical protein was developed by Woese (1965) and the concept of frozen codon assignments was pointed out by Crick (1968). Knight et al. (1999) argued that three different signals are present in the stucture of the genetic code table, which they called the "faces of the genetic code". The chemical face of the code is the effect of UAUUGGGG, the leucine, phenylalanine, tyrosine, tryptophan and histidine anticodons, and the arginine aptamers. The historical face of the code is the effect of tRNA$^{Asn}$ evolving from tRNA$^{Asp}$, tRNA$^{Gln}$ evolving from tRNA$^{Glu}$, tRNA$^{Cys}$ evolving from tRNA$^{Ser}$,

tRNA$^{Lys}$ evolving from tRNA$^{Arg}$, tRNA$^{Met}$ evolving from tRNA$^{Thr}$, and so on. But how came the selection face to be?

Often, the realization that "the genetic code is one in a million" (Freeland and Hurst, 1998) is immediately followed by the assumption that myriads of codes have to be screened by natural selection to evolve such a characteristic. Ardell (1998) pointed out that related variants of the genetic code, could compete on error minimization, and thus a much more restricted number of codes could suffice for code optimization, a concept he called "the sequential model for evolution of the code by error minimization" (Ardell, 1998). This concept is logically equivalent to the view of the primary emergence of short protein sequences, stressed by de Duve (2005). Not all possible mRNA sequences were tried, the first proteins were small. Not all possible genetic codes were tried, tRNA space was invaded in the same way as protein space. Most of the space was never visited. De Duve (2005) described this as follows: "An important implication of the necessarily stepwise course of RNA and protein lengthening is that exploration of the corresponding sequence spaces was itself stepwise" (de Duve, 2005).

With the order of codon assignments presented in the AR-NE-GRERO model in mind, we can now scrutinize how the error robustness came to be. The 4 tRNA genetic code already showed the basics of the error robustness: the polar requirements of the columns were in principle: 5, 7, 13 and 8. The phase of ambiguity reduction was responsible for error robustness in the first position of the codon for the left half of the table, and in the third position of the codon for the GAN box. So, error robustness is a reflection of the statistical nature of the primitive proteins, for this part of the error robustness. The assignment of GUC to valine was a crucial event, enabling the ambiguity reduction by using the aptamer characteristics of UAUUGGGG and GAG, and enabling later expansion of the code, without losing error robustness, by using the aptamer characteristics of CAA and GAA. Choosing GUC to be valine was like buying the winning number in a lottery, a choice which in evolutionary biology is called a preadaptation. Everybody walking a different way was bound to get lost in the storm of competition.

The 8 tRNA genetic code showed 4.9. 4.9 and 5.6 in the first column, 6.6/7.5 and 7.0 in the second column, 12.5 and 13.0 in the GAN box and 7.9 for the GGY codons. The phase of nonsense elimination was responsible for an enormous amount of error robustness in the third position of the codon. CUN, GUN, ACN, GCN and GGN became family boxes, and stayed that way. Later, in the next phase, also the UCN, CCN and CGN boxes would become family boxes. The leading themes were nonsense elimination and the remaining part of ambiguity reduction, error robustness was a by-product. But this by-product made the mRNA/protein system incredibly powerful, both in error robustness in an age of still-poor fidelity, and in evolvability of new biocatalysts (see e.g. Wagner, 2005). With the emergence of the 6 arginine codons (using the aptamer characteristics of these codons), also the assignment of glycine to GGN (which would be a chemical determined assignment itself, if Copley et al., 2005, are right in their idea's on amino acid synthesis from keto acids by catalytic dinucleotides) turns out to be a preadaptation, because 7.9 and 9.1 are rather close in polar requirement.

The assignments which make the difference in being a little error robust or being incredibly error robust are: phenylalanine (5.0) and leucine (4.9) sharing a box, cysteine (4.8 or 5.5) and tryptophan (5.2 or 5.3) sharing a box, histidine (8.4) and glutamine (8.6) sharing a box, isoleucine (4.9) and methionine (5.3) sharing a box and asparagine (10.0) and lysine (10.1) sharing a box. The first of these is chemically determined, but in the remaining 4 cases one of the two amino acids could have been an amino acid with a totally different polar requirement (in that case, an "accident" would have been frozen). We thus find in the third phase a strong selective pressure on error robustness in polar requirement in the third position of the codon. This pressure is so high that it is unacceptable to attribute it to error robustness in protein structure, but an alternative is present in the need of highly specific interaction with cofactor nucleotides. Thus, the original function of polar requirement as a measure for interactions between amino acid and nucleic acid (Woese et al., 1966) is back.

What about error robustness in the second position? Error robustness because histidine (8.4) and arginine (9.1) are in the same row, is a by-product of the aptamer characteristics of arginine codons and of anticodons decoding histidine: here we are looking to the chemical face of the genetic code, not the selection face. Error robustness because lysine (10.1) and arginine (9.1) are in the same row, is a byproduct of the potential of a tRNA$^{Arg}$ to develop into a tRNA$^{Lys}$: here we are looking to the historical face of the genetic code. Error robustness because leucine (4.9) and proline (6.6) are in the same row, is a by-product of the ambiguity reduction processes in the first and the second column: we see two of the three faces, the chemical face (GAG necessarily decodes leucine) and the historical face (tRNA$^{Pro}$ was a result of duplication of the primordial tRNA$^{Thr}$, followed by subfunctionalization), but when we think we see the selection face, we are mistaken. I cannot find a case in which second position error robustness clearly was the main driving force in an assignment. Thus Ardell (1998) seems to be mistaken in the conclusion that second position transitions were selected to be error robust on the transcription level, exactly as Judson and Haydon (1999) argue.

## *Concluding Comments*

If the AR-NE-GRERO model is the right one, then Judson and Haydon (1999) were also right in another conclusion they arrive at. The main ultimate result of the process of evolution of the genetic code was: evolvability (Speyer et al., 1963: "Two selective advantages may accrue from the extensive degeneracy of the code; mutational nucleotide changes will result in a wider range of amino acid replacements, and in a less frequent production of nonsense triplets"; Maeshiro and Kimura, 1998; Judson and Haydon, 1999; Freeland, 2002; Wagner, 2005). But according to the AR-NE-GRERO model, error robustness optimization *was* a selective factor, especially in the assignments of glutamine and lysine.

One of the strongest facts supporting the model should be mentioned once again. The model (uniquely) explains the mirror structure of the two aminoacyl-tRNA synthetase phylogenetic trees: Leu/Val/IleRS facing Pro/Thr/SerRS, MetRS facing HisRS, CysRS facing GlyRS, ArgRS facing AlaRS, GluRS facing AspRS, LysRS-I facing LysRS-II, TyrRS facing PheRS, and TrpRS facing SepRS.

**Acknowledgements**

For discussion, encouragement and sending me literature, I thank Wouter Hoff, Wander Sprenger, Kees Breek, Dave Speijer, Christian de Duve, Carl Woese, Damien Matthew, Harry Buhrman, Leen Stougie, Steven Kelk, Wouter Koolen and Steven de Rooij. For hospitality, I thank the Centrum voor Wiskunde en Informatica.